\documentclass[letter]{aa}
\usepackage[varg]{txfonts}
\usepackage{natbib}
\usepackage{graphicx}
\bibpunct{(}{)}{;}{a}{}{,} 
\begin{document}
\title{Examining the T Tauri system with SPHERE
\thanks{Based on observations made at the La Silla Paranal Observatory under programme ID 60.A-9363(A) and 60.A-9364(A).}}
\subtitle{}
\author{
Gergely~Cs\'ep\'any\inst{1,2}
\and
Mario~van~den~Ancker\inst{1}
\and
P\'eter~\'Abrah\'am\inst{2}
\and
Wolfgang~Brandner\inst{3}
\and
Felix~Hormuth\inst{3}
} 
\institute{European Southern Observatory, Karl-Schwarzschild-Str. 2, 85748 Garching bei M\"unchen, Germany
\and 
Konkoly Observatory of the Research Centre for Astronomy and Earth Sciences, Hungarian Academy of Sciences, Konkoly Thege Mikl\'os \'ut 15-17, 1121 Budapest, Hungary
\and 
Max-Planck-Institut für Astronomie,  K\"onigstuhl 17,  69117 Heidelberg, Germany}
\date{Received / Accepted }
\abstract{
The prototypical low-mass young stellar object, T Tauri, is a well-studied multiple system with at least three components. 
}
{
We aim to explore the T~Tau system with the highest spatial resolution, study the time evolution of the known components, and re-determine the orbital parameters of the stars.}
{
Near-infrared classical imaging and integral field spectrograph observations were obtained during the Science Verification of SPHERE, the new high-contrast imaging facility at the VLT. 
The obtained FWHM of the primary star varies between 0.050\arcsec\ and 0.059\arcsec, making these the highest spatial resolution near-infrared images of the T~Tauri system obtained to date.}
{
Our near-infrared images confirm the presence of extended emission south of T~Tau~Sa, reported in the literature. 
New narrow-band images show, for the first time, that this feature shows strong emission in both the Br-$\gamma$ and H$_2$ 1-0 S(1) lines. Broadband imaging at 2.27~$\mu$m shows that T~Tau~Sa is 0.92~mag brighter than T~Tau~Sb, which is in contrast to observations from Jan. 2014 (when T~Tau~Sa was fainter than Sb), and demonstrates that T~Tau~Sa has entered a new period of high variability. 
The newly obtained astrometric positions of T~Tau~Sa and Sb agree with orbital fits from previous works. 
The orbit of T~Tau~S (the center of gravity of Sa and Sb) around T~Tau~N is poorly constrained by the available observations and can be fit with a range of orbits ranging from a nearly circular orbit with a period of 475 years to highly eccentric orbits with periods up to $2.7 \times 10^4$ years.
We also detected a feature south of T~Tau~N, at a distance of $144\pm 3$~mas, which shows the properties of a new companion.
}
{}
\keywords{Stars: variables: T Tauri, Herbig Ae/Be - stars: pre-main sequence - stars: evolution - binaries: visual - techniques: high angular resolution - instrumentation: adaptive optics}
\maketitle

\section{Introduction}

T Tauri (HD~284419), initially considered to be the prototype of 
a class of low-mass pre-main sequence stars, is now known to be 
a complex young system of at least three stellar components (T~Tau~N, 
and the close binary system T~Tau~Sa/Sb) as well as of jets, and 
outflows. Whereas T~Tau~N suffers little extinction \citep[$A_V$ = 1\fm46;][] 
{2001ApJ...556..265W}, T~Tau~Sb is hidden behind about 15 mag of 
extinction, while T~Tau~Sa appears to be even more obscured 
\citep{1999A&A...348..877V, 2010A&A...517A..16V, 2005ApJ...628..832D}. 

Orbital motion of Sa and Sb has been measured over the last 16 years, 
yielding a well-constrained orbital model for Sa-Sb with a semi-major 
axis of 89~mas and a period of 29~years \citep{2014AJ....147..157S}. 
Using the orbital data, the mass of T~Tau~Sa was shown to be 1.9--2.3~M$_\sun$ 
and the mass of Sb is between 0.7 and 0.9 M$_\sun$ \citep{2006A&A...457L...9D, 
2008A&A...482..929K}. The orbital motion of the Sa-Sb system around T~Tau~N 
is less well constrained, with initial fits yielding an orbital periods 
between 475 and 23000 years, and a total system mass between 
4.5 and 5.6 M$_\sun$ \citep{2004AJ....128..822J, 2008A&A...482..929K}.

Millimeter interferometry first revealed the 
presence of a circumstellar disk 
around the optically dominant component T~Tau~N 
\citep{1997ApJ...490L..99H, 1998ApJ...505..358A}. The presence 
of CO absorption in the spectrum of T~Tau~Sa suggests the presence
of a circumstellar disk viewed nearly edge-on around this source as 
well \citep{2005ApJ...628..832D}. 
In the high-resolution images \citet{2010A&A...517A..16V} found 
circumstellar emission around T~Tau~Sa in the $K$-band on scales of 
5 to 20--30 AU. It is as of yet unclear whether this emission can be 
directly associated with the tentative edge-on disk of T~Tau~Sa. 
All three components of the T Tau system are actively accreting and 
are spatially unresolved hydrogen recombination line emitters 
\citep{2002ApJ...568..267K, 2010A&A...517A..19G}.

In this {\em paper} we present new infrared images and spectroscopy of the T Tau system, 
obtained with SPHERE at the VLT. At small scales, these new images present 
the highest contrast data obtained so far of T Tau. We discuss the orbital 
and photometric evolution of the known components of T Tau and 
present evidence for the presence of a new tentative companion south of T~Tau~N.

\section{Observations and Data Analysis}

\begin{figure}
\centering
\includegraphics[width=0.45\hsize]{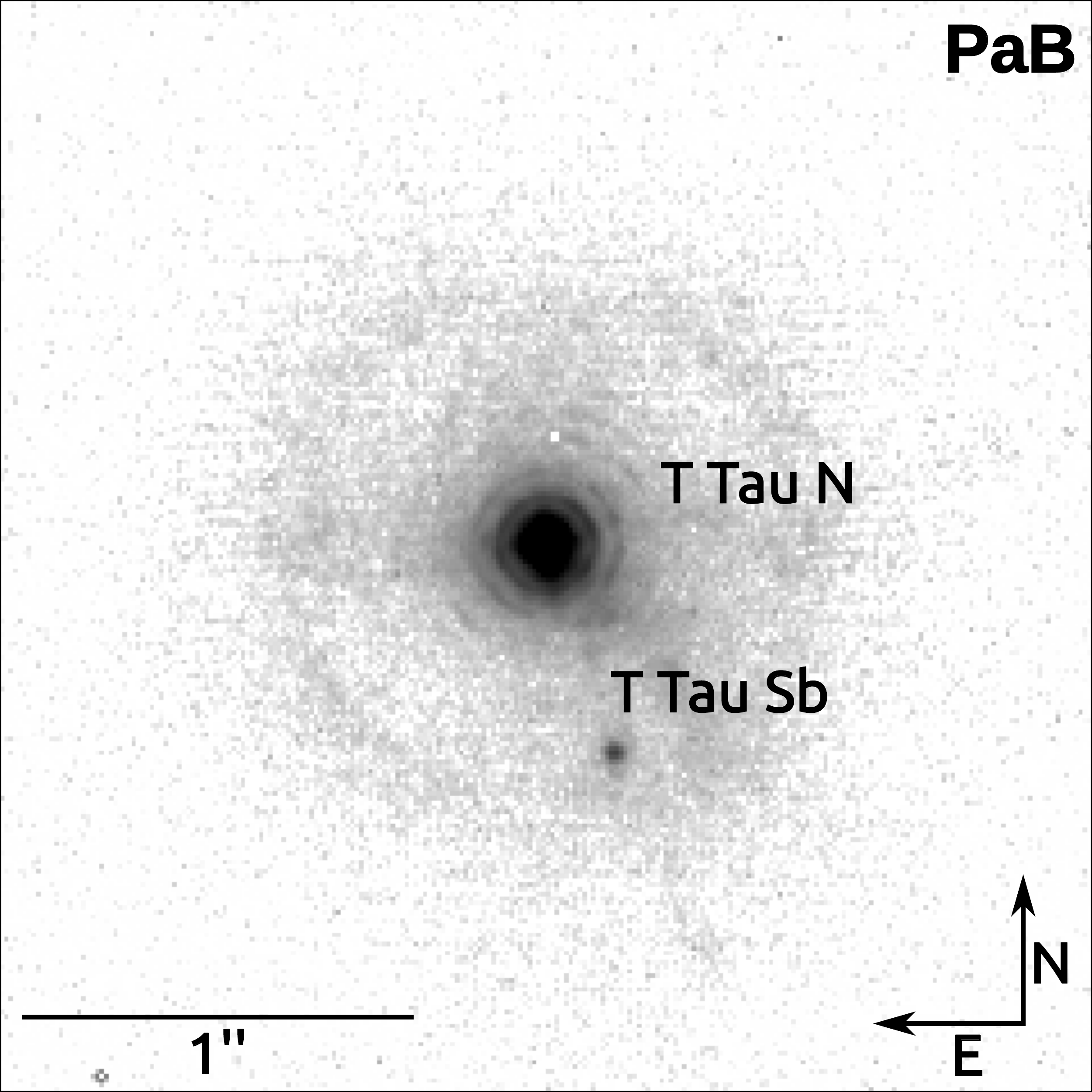}
\includegraphics[width=0.45\hsize]{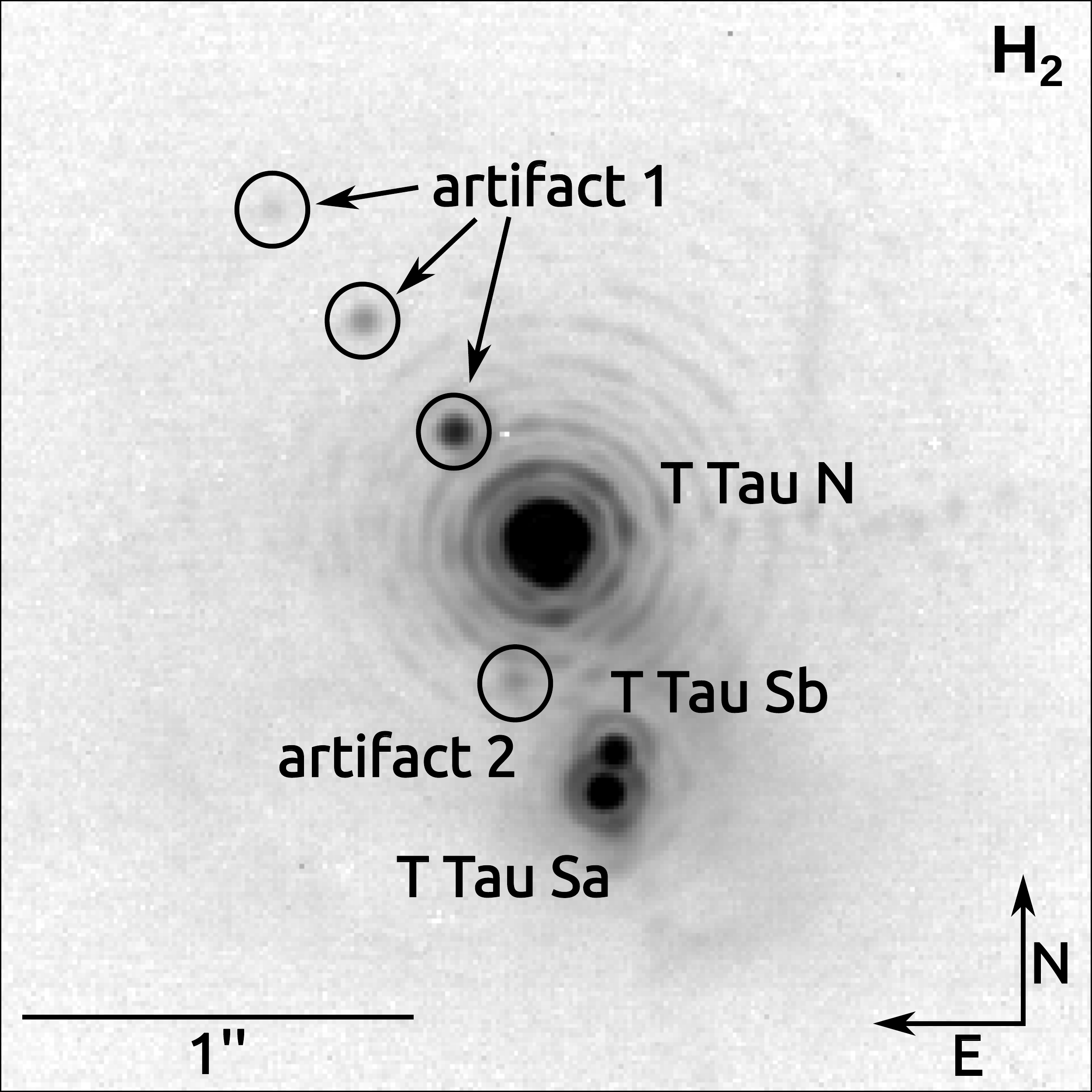}
\includegraphics[width=0.45\hsize]{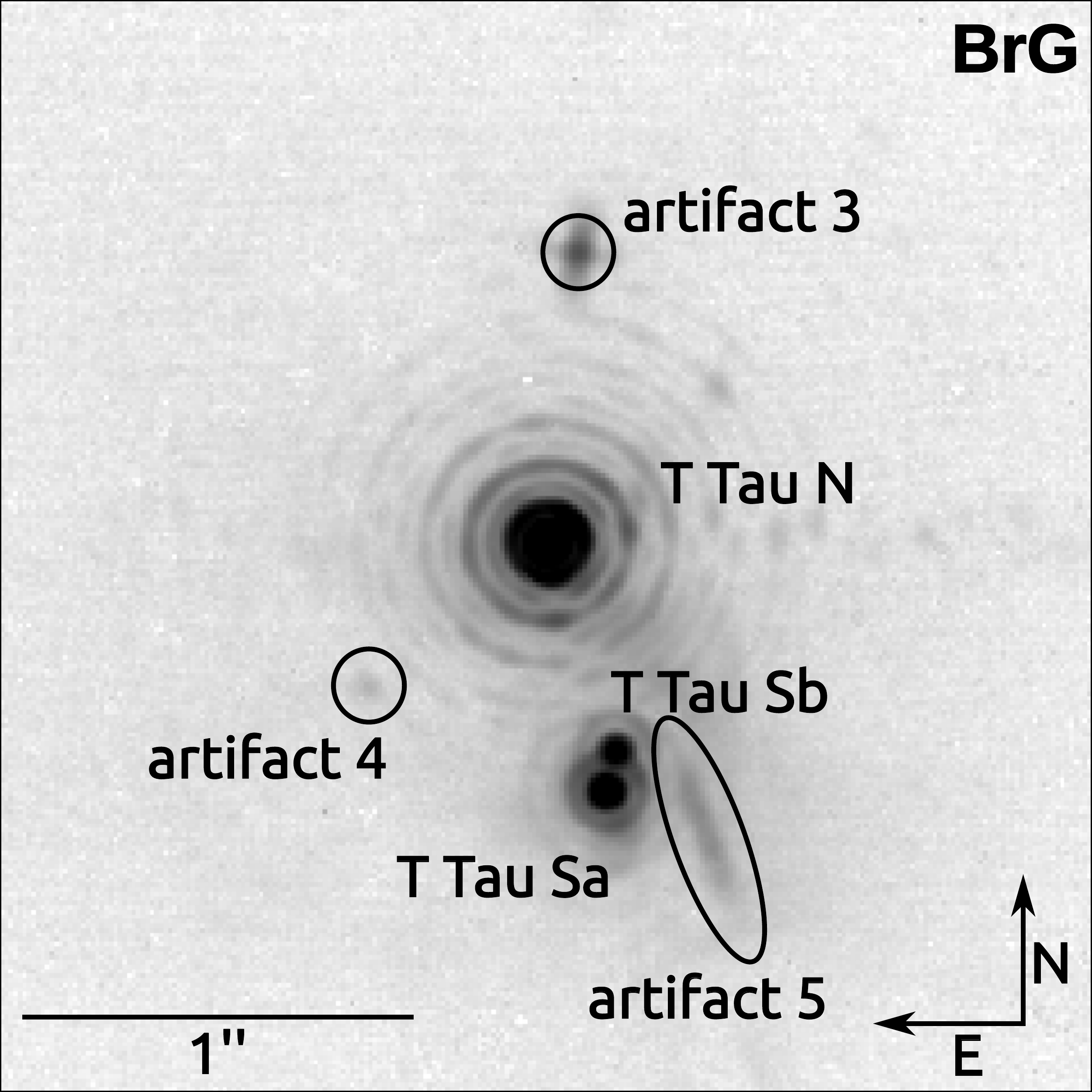}
\includegraphics[width=0.45\hsize]{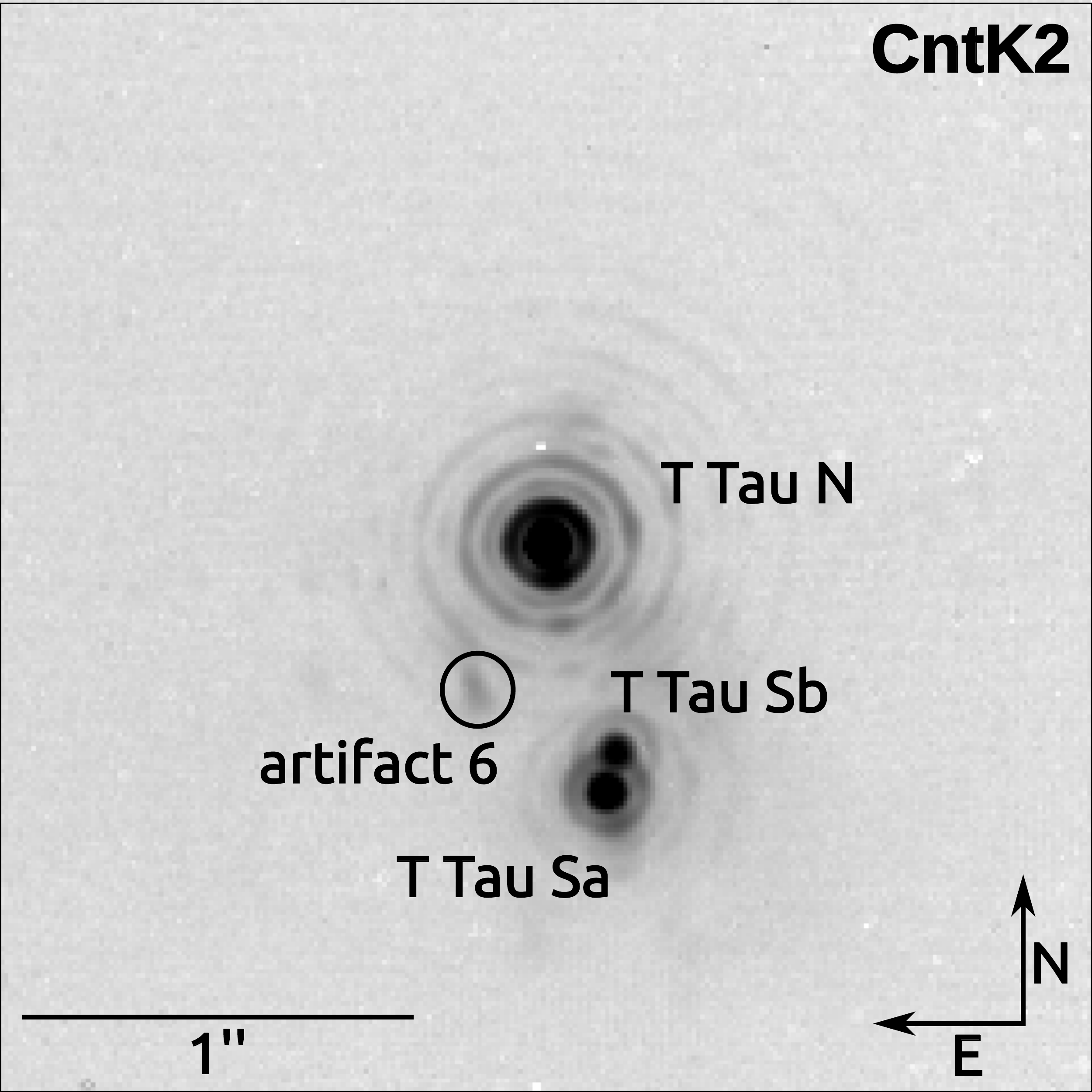}
\caption{Classical images of the T Tauri system in different filters with logarithmic intensity scaling. From left top to bottom right: 
Pa-$\beta$ ($1.281\ \mathrm{\mu m}$), 
H$_2$ ($2.128\ \mathrm{\mu m}$), 
Br-$\gamma$ ($2.163\ \mathrm{\mu m}$), 
and 
CntK2 ($2.270\ \mathrm{\mu m}$).
}
\label{fig:motions}
\end{figure}

Observations of the T Tau system were acquired during the Science Verification of SPHERE\footnote{http://www.eso.org/sci/facilities/paranal/instruments/sphere.html}, the newly installed extreme Adaptive Optics facility at the VLT. The first set of observations was obtained on Dec. 9$^\mathrm{th}$, 2014, when SPHERE was operated in the IRDIS classical imaging mode with the ContJ ($1.216\ \mathrm{\mu m}$), Pa-$\beta$ ($1.281\ \mathrm{\mu m}$), H$_2$ ($2.128\ \mathrm{\mu m}$), Br-$\gamma$ ($2.163\ \mathrm{\mu m}$) and CntK2 ($2.270\ \mathrm{\mu m}$) filters. 
Calibration frames were acquired for the geometrical distortion map, which were used during the data processing to ensure astrometric precision. A second set of observations was acquired on Jan. 22$^\mathrm{nd}$, 2015, when SPHERE was operated in the IRDIFS mode with a 155 milli-arcsecond (mas) diameter apodized Lyot coronagraph, offering simultaneous Integral Field Spectrograph (IFS) observations from 0.95--1.65~$\mu$m and imaging in the Ks ($2.181\ \mathrm{\mu m}$) filter. Observations with the star offset from the coronagraphic mask were also obtained in order to be able to measure total fluxes. 
A PSF reference star (TYC 1290-457-1) was observed right after the IRDIFS observation of T Tau, at similar airmass and atmospheric conditions as the science target.

The obtained data sets were reduced with the pre-release version 0.14.0-2 of the SPHERE pipeline. Apart from the usual steps of bias subtraction, flat-fielding and wavelength calibration, the pipeline also includes a correction for geometric distortions. Post-processing of the data was done using the FITSH software package \citep{2012MNRAS.421.1825P} to sum the data resulting from the two IRDIS channels. The achieved FWHM of the diffraction-limited core of T~Tau~N varies between 0.050\arcsec\ and 0.059\arcsec\ (from ContJ to CntK2), making these the highest spatial resolution near-infrared images of the T Tau system obtained to date. The classical images are shown in Fig.~\ref{fig:motions}. The H$_2$, Br-$\gamma$ and CntK2 filters suffer from a number of known ghosts which can be seen close to bright stars. 
Figure~\ref{fig:ifsplots} features the IFS images taken with the coronagraph, in which a cross-like pattern can be seen that is caused by reflections from the wires on which the coronagraphic mask is suspended. 
The longer wavelength frames in the IFS observation suffer from a flat fielding error pattern; either the flat fields were inadequate or the pre-release version of the pipeline failed to properly reduce the data. We omit performing precise astrometry and photometry in those frames, and only use them to determine the spectrum with a larger aperture to compensate for the error pattern.

Positions and fluxes (both relative to T~Tau~N) of the stellar components T Tau Sa and Sb were 
measured and are listed in Table~\ref{table:obs}. Relative fluxes in this Table were computed 
using aperture photometry, with the aperture radius taken equal to the radius of the Airy disc 
(73.5 mas). The listed uncertainties of the stellar positions 
and angles include the error of the centroid fitting, the pixel-scale conversion uncertainty 
and relative atmospheric diffraction. The latter error contribution is calculated by assuming 
the maximum relative shift in position of the components diffraction in the used filter.

\section{Results}
\subsection{Classical Imaging}
\begin{table}
\setlength{\tabcolsep}{3.5pt}
\caption{\label{table:obs}Positions and magnitude differences (relative to T Tauri N) of 
the stellar components.}
\centering
\begin{tabular}{lllll}
\hline\hline
Object & Filter & Sep. (\begin{tiny}mas\end{tiny}) & PA ($^\circ$) & Magn. diff. \\
\hline
T~Tau~Sa        & H$_2$ & $\phantom{m}688 \pm   3$ & $ 193.4 \pm 0.2 $ & $ 2.41 \pm 0.01 $ \\
                & Br-$\gamma$ & $\phantom{m}686 \pm   3$ & $ 193.4 \pm 0.2 $ & $ 2.26 \pm 0.01 $ \\
                & CntK2 & $\phantom{m}685 \pm   3$ & $ 193.3 \pm 0.2 $ & $ 1.95 \pm 0.01 $ \\
                & Ks & $\phantom{m}687 \pm   15$ & $ 193.1 \pm 1.0 $ & $ 1.82 \pm 0.03 $ \\
T~Tau~Sb        & ContJ & $\phantom{m}589 \pm   6$ & $ 198.2 \pm 0.4 $ & $ 6.14 \pm 0.12 $ \\
                & Pa-$\beta$ & $\phantom{m}590 \pm   6$ & $ 198.2 \pm 0.4 $ & $ 5.77 \pm 0.09 $ \\
                & H${}_2$ & $\phantom{m}595 \pm   3$ & $ 197.9 \pm 0.3 $ & $ 3.15 \pm 0.01 $ \\
                & Br-$\gamma$ & $\phantom{m}593 \pm   3$ & $ 198.0 \pm 0.3 $ & $ 3.09 \pm 0.02 $ \\
                & CntK2 & $\phantom{m}592 \pm   3$ & $ 197.9 \pm 0.3 $ & $ 2.87 \pm 0.02 $ \\
                & Ks & $\phantom{m}589 \pm   15$ & $ 197.7 \pm 1.3 $ & $ 2.58 \pm 0.05 $ \\
Tentative       & 1.02~$\mu$m & $140.9 \pm  2.7 $ & $ 199.0 \pm   2.8 $ & $ 4.96 \pm 0.30 $ \\
companion       & 1.19~$\mu$m & $143.7 \pm  2.5 $ & $ 198.0 \pm   1.6 $ & $ 4.39 \pm 0.19 $ \\
(IFS obs.)      & 1.34~$\mu$m & $145.8 \pm  3.1 $ & $ 197.2 \pm   2.0 $ & $ 4.37 \pm 0.30 $ \\
\hline
\end{tabular}
\end{table}

The two shortest wavelength images (ContJ and Pa-$\beta$) show the primary (T~Tau~N) and one 
of the southern components (T~Tau~Sb). In addition, some extended emission may be present at 
low intensities. In the longer wavelength images (H$_2$, Br-$\gamma$ and CntK2), the second 
known southern component (T~Tau~Sa) is also visible. In all images, the stellar sources 
T~Tau~N, Sa and Sb appear indistinguishable from a point-source. 

Observational $3 \sigma$ detection limits in the different filters were calculated from our 
data and are shown in Fig.~\ref{fig:obslimits} (online material) by calculating the magnitude of a hypothetical source that is hidden in the background noise at the given location. The bumps in the plots at $\approx 250$ 
and $\approx 600$ mas are the result of the Airy rings. We do not detect the tentative 
northern component, denoted as T Tau (O') by \citet{1985ApJ...297L..17N} 
\citep[and later also observed by][]{1991A&A...249..392M}. Our upper limits show that 
this tentative component, if exists, must be at least 4.7 magnitudes fainter 
than T Tau N over the entire 1.2--2.2~$\mu$m range covered by our observations.

In our CntK2 ($2.27\ \mathrm{\mu m}$) image, T~Tau~Sa appears distinctly brighter (0.92 mag, 
Table~\ref{table:obs}) than T~Tau~Sb. This is in marked contrast to the latest $K$-band brightness 
ratio measurements, obtained in January 2014, in which Sa was about 0.8 magnitudes fainter than 
Sb \citep{2014AJ....147..157S}. This variability of T~Tau~Sa of 1.7 magnitudes in less than a year 
is similar to the photometric behaviour shown prior to 2006 \citep{2004ApJ...614..235B, 2005ApJ...628..832D}, 
and demonstrates that the period of lesser variability reported from 2006--2013 
\citep{2010A&A...517A..16V, 2014AJ....147..157S} has come to an end.

In addition to the continuum filters, our data set includes measurements in several narrow-band filters 
centered on well-known emission lines (Pa-$\beta$, Br-$\gamma$ and the H$_2$ 1-0 S(1) line). From the 
brightness ratio of Br-$\gamma$ to the nearby CntK2 filter we conclude that T~Tau~N (brightness ratio 
1.52) shows strong Br-$\gamma$ emission, most likely linked to the accretion onto the central star. 
Measurements for Sa and Sb (brightness ratios of 1.14 and 1.22 respectively) are inconclusive, 
although we note that Br-$\gamma$ emission has previously been detected from both these components 
\citep{2010A&A...517A..19G}.

For H$_2$, the brightness ratio H$2$/CntK2 is very high (1.90) for T~Tau~N, 
demonstrating the presence of strong H$_2$ 1-0 S(1) emission close to this component. 
Previous studies found H$_2$ emission around T~Tau~N and S \citep{2002ApJ...568..267K, 2004ApJ...614..235B, 2010A&A...517A..19G}, which our observations also confirm.
Our CntK2, Br-$\gamma$ and H$_2$ images (Fig.~\ref{fig:cntk2_subtr} in the online material) show  extended emission at a peak flux of $6 \sigma$ above background level south of T Tau Sa, first described 
by \citet[][ Fig.~\ref{fig:cntk2_subtr}]{2010A&A...517A..16V}. 
The ratio\,of\,the\,flux\,in\,the H$_2$ and 
Br-$\gamma$ filters to that in CntK2 (1.78 and 1.53, respectively) is high, demonstrating that this 
feature possesses emission in the Br-$\gamma$ and the H$_2$ 1-0 S(1) lines.

\subsection{Integral Field Spectrograph}

\begin{figure}
\centering
\includegraphics[width=0.49\hsize]{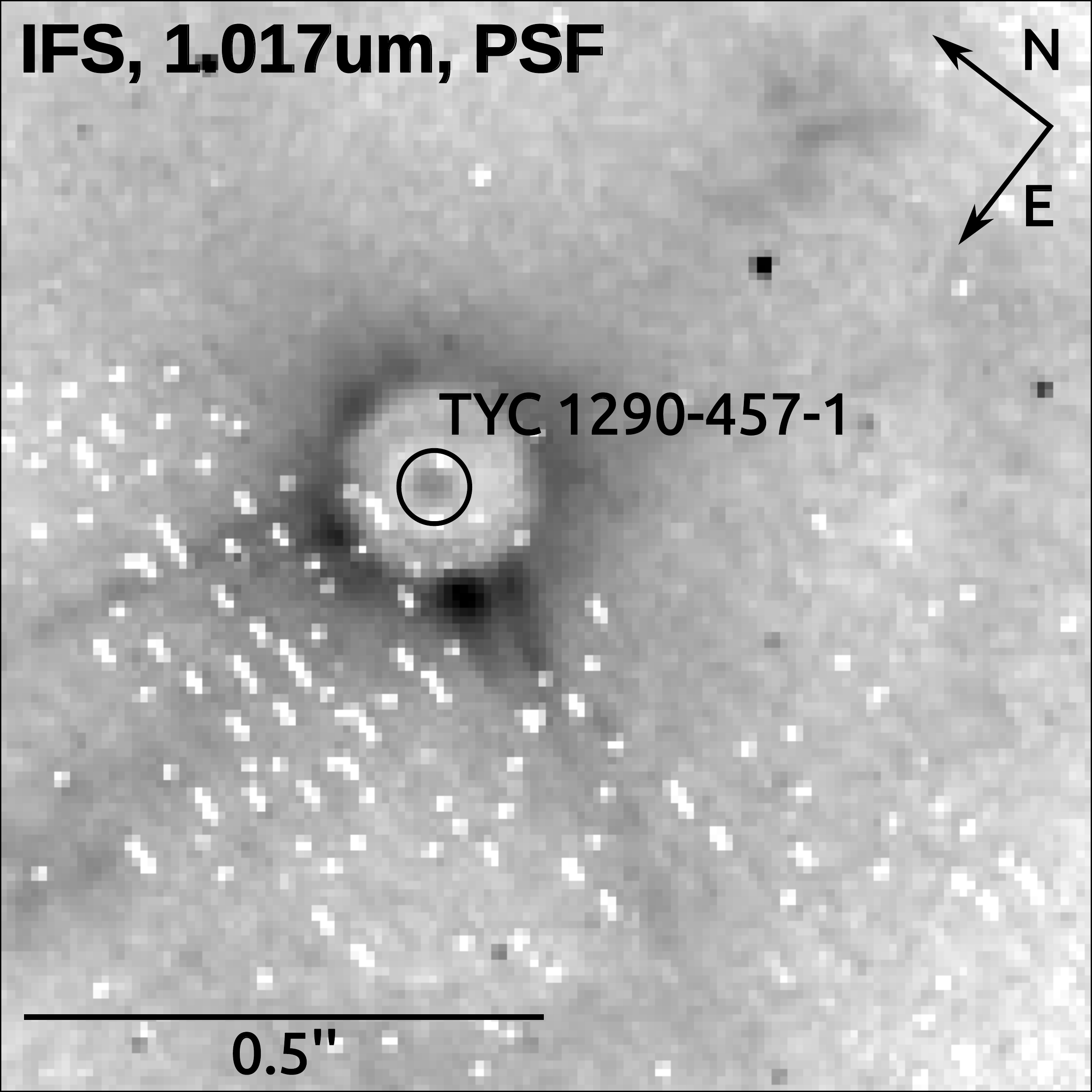}
\includegraphics[width=0.49\hsize]{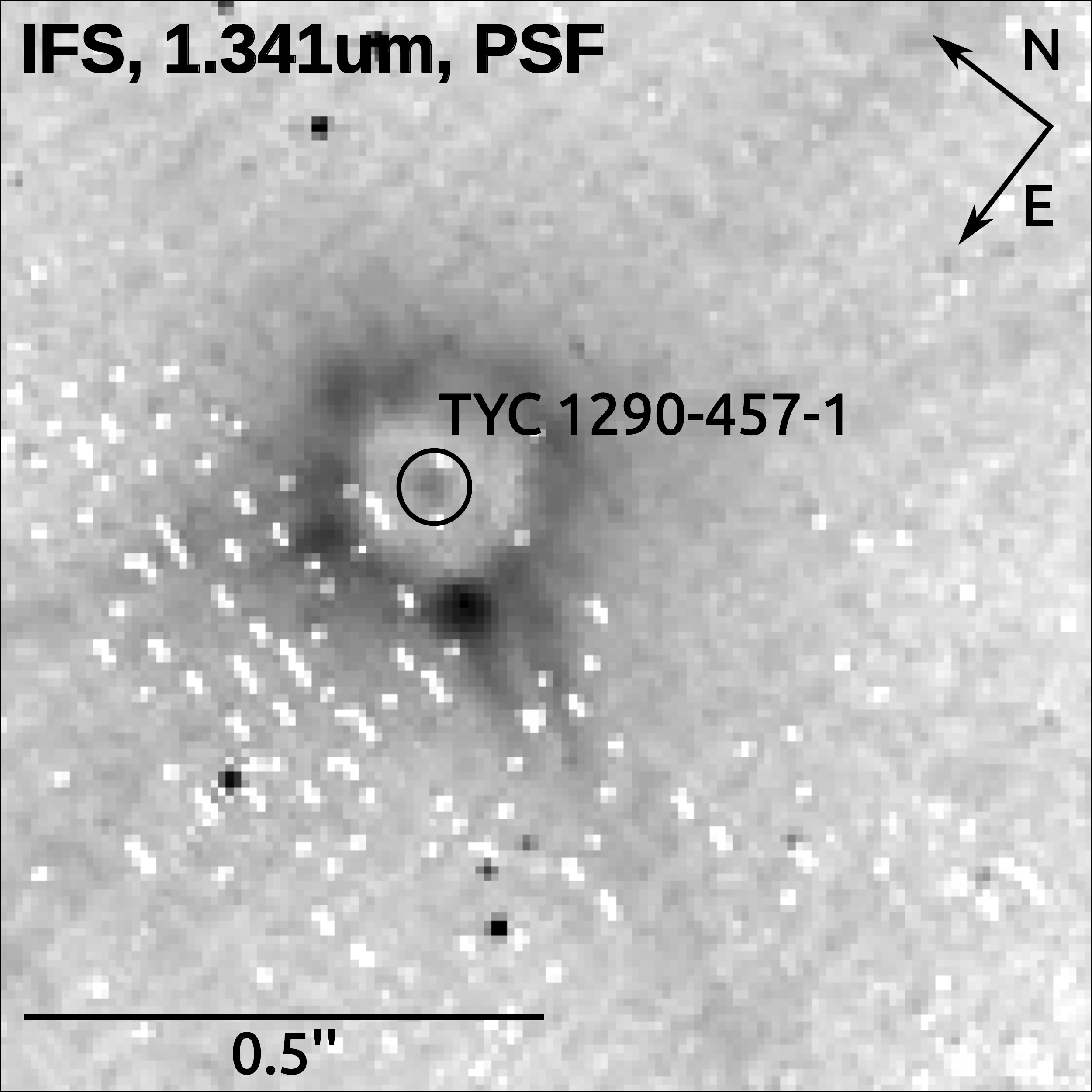}
\includegraphics[width=0.49\hsize]{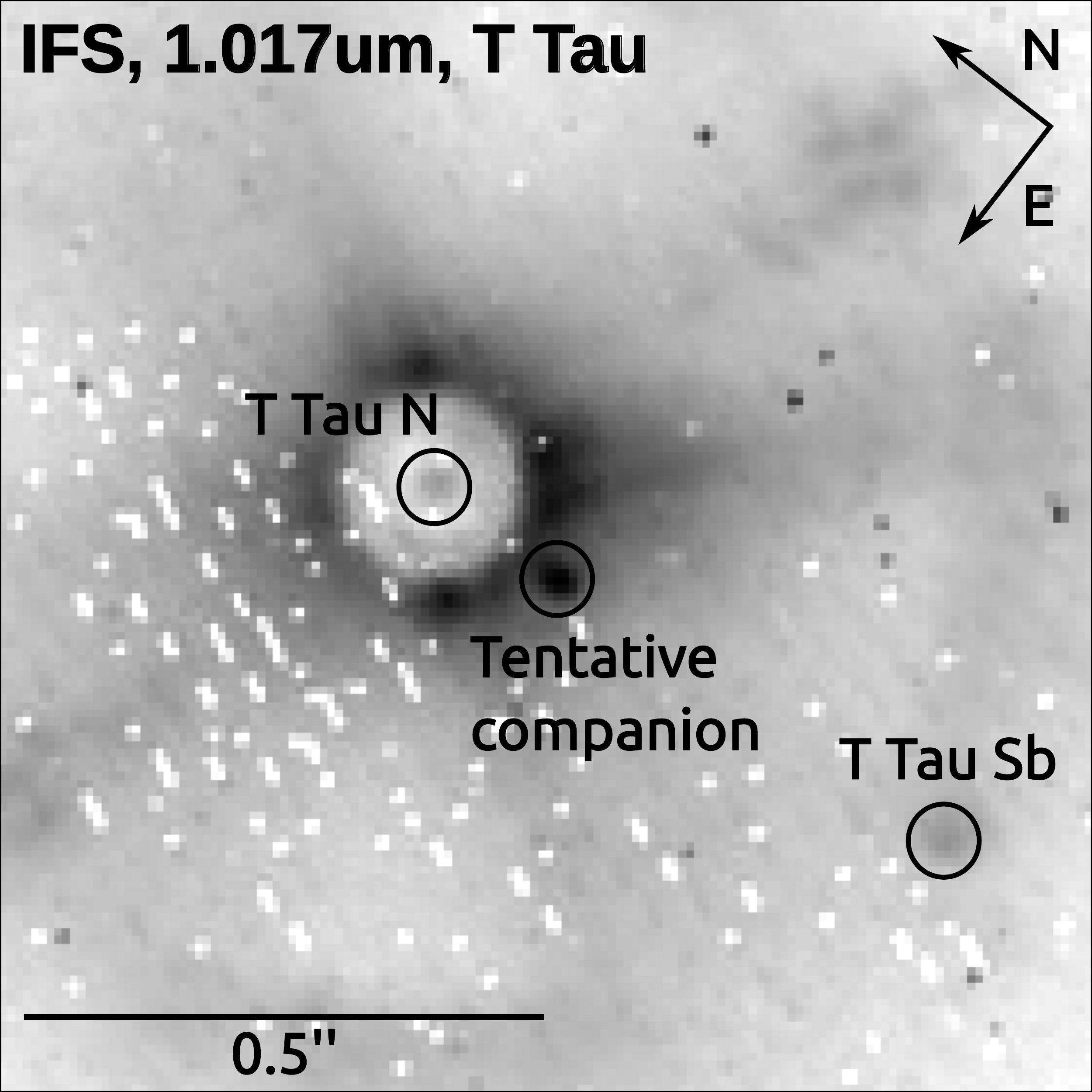}
\includegraphics[width=0.49\hsize]{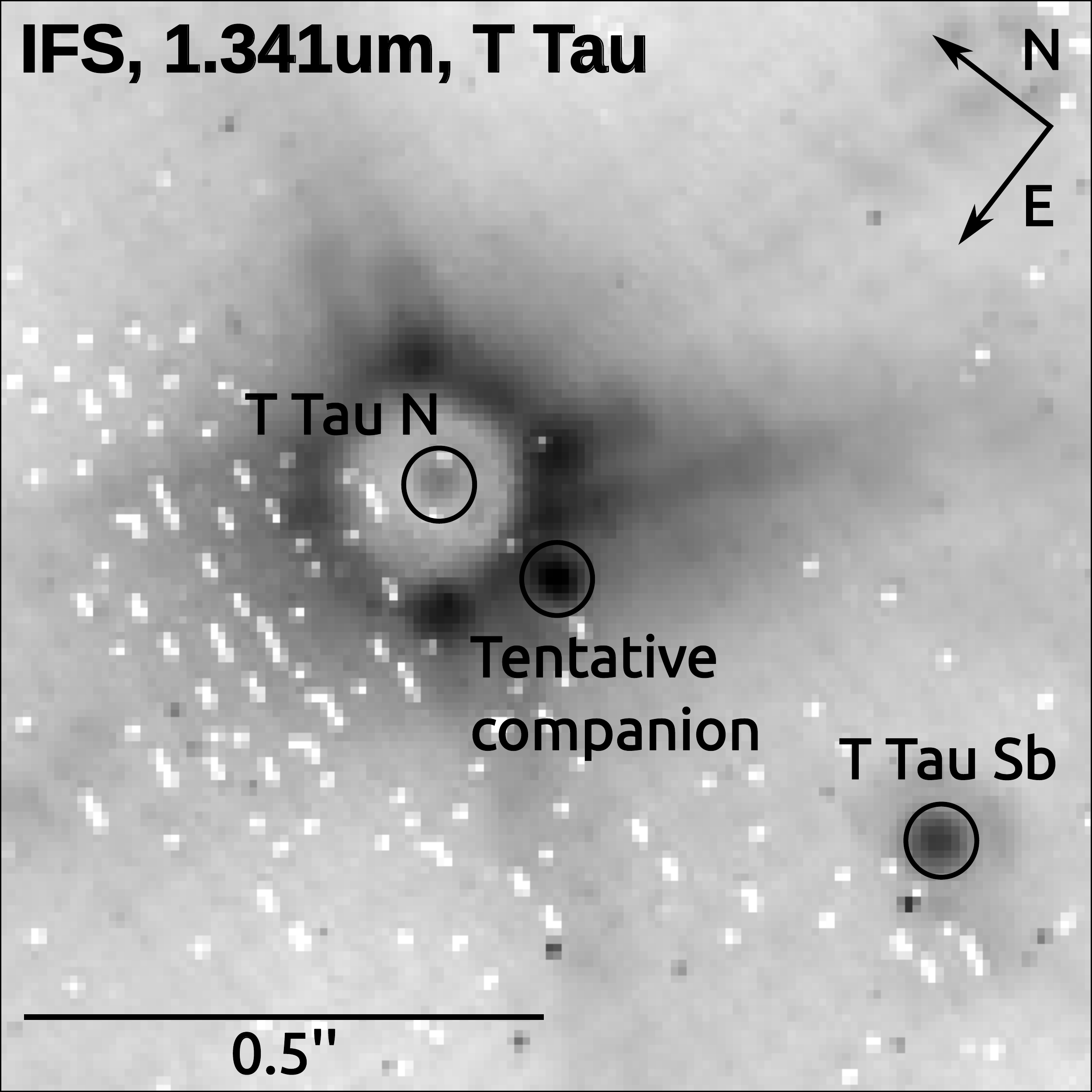}
\includegraphics[width=0.49\hsize]{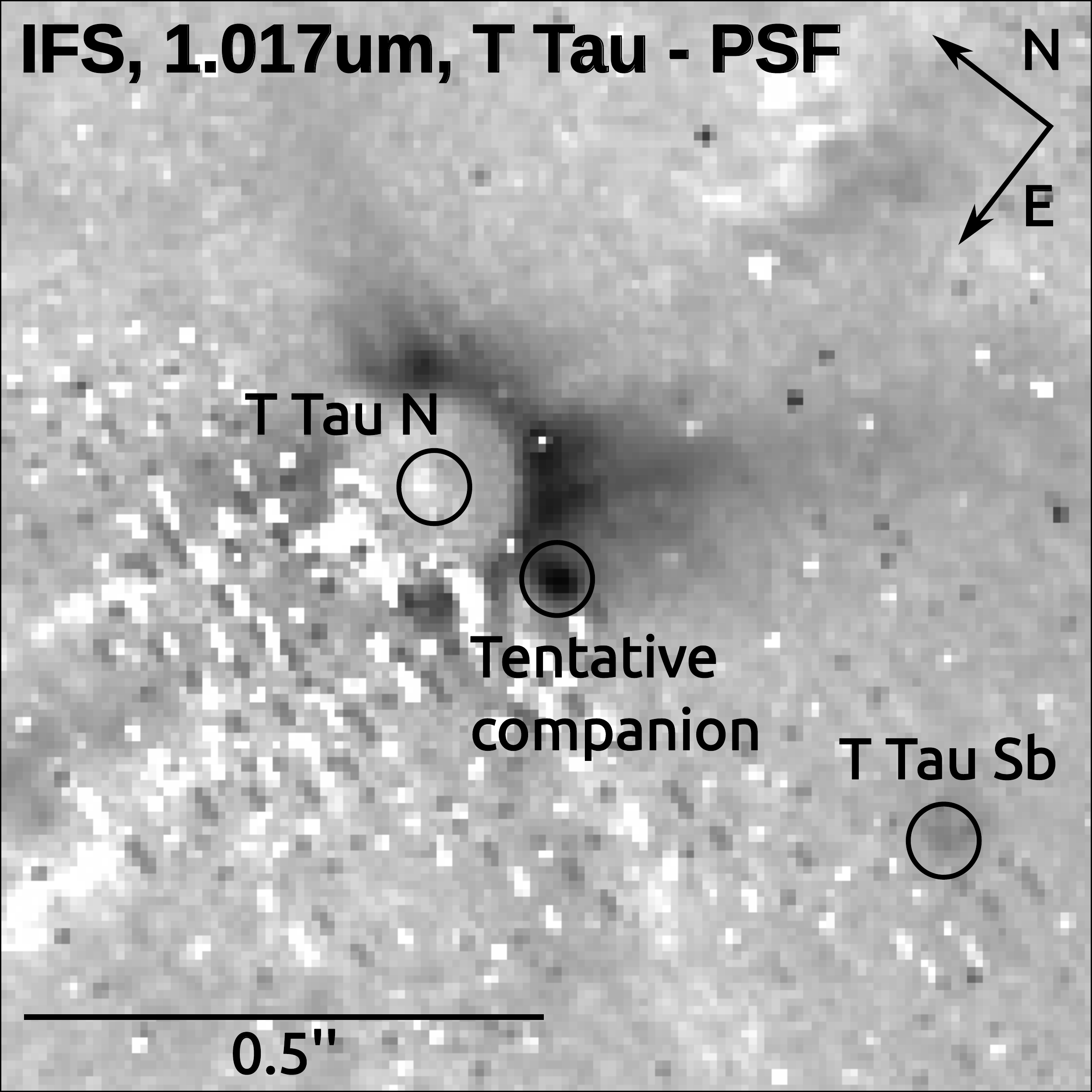}
\includegraphics[width=0.49\hsize]{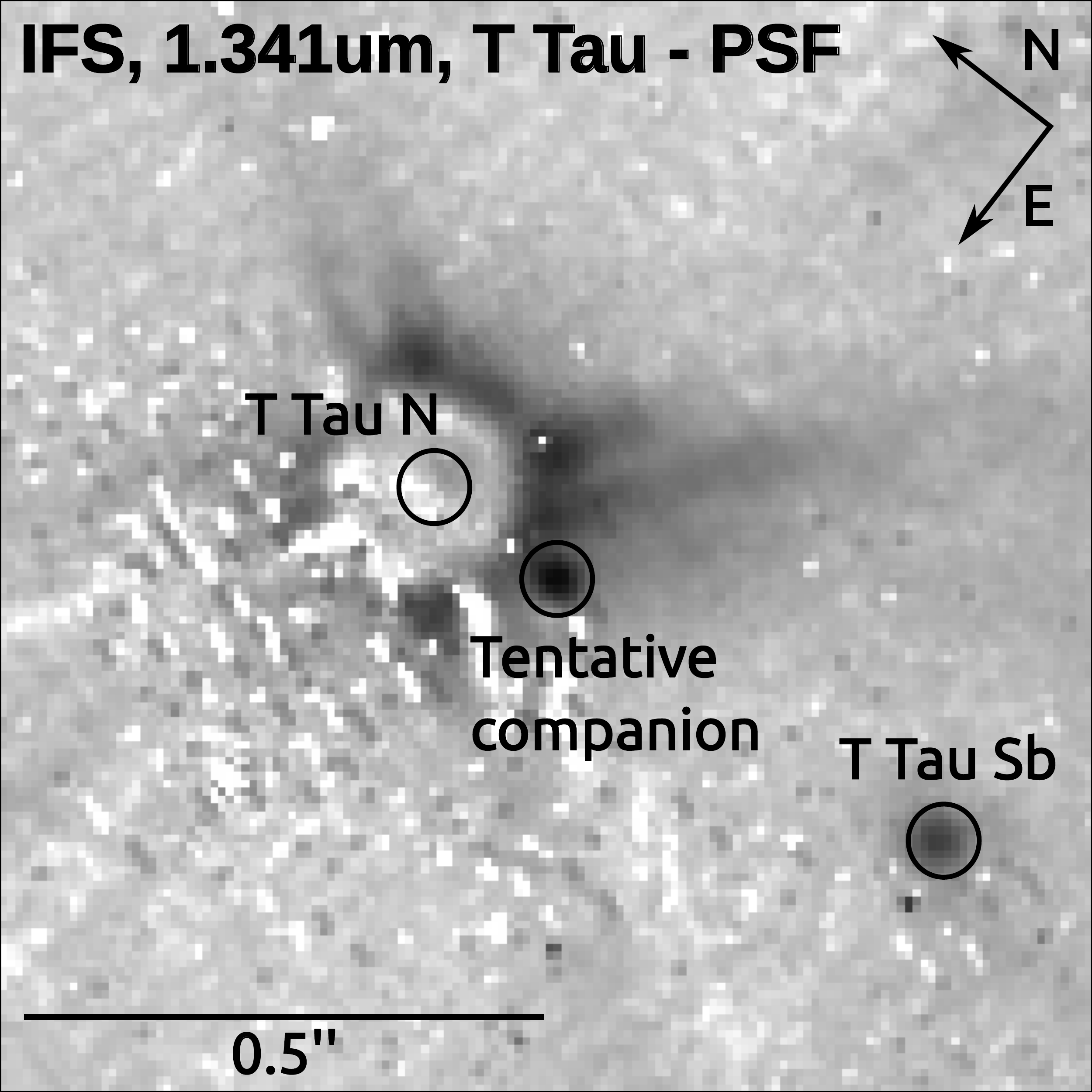}
\caption{The T Tauri system as seen at two different wavelengths in the IFS observation. \textit{Top:} PSF reference star (TYC 1290-457-1), \textit{Middle:} T~Tauri system, \textit{Bottom:} T Tauri with the PSF reference subtracted.}
\label{fig:ifsplots}
\end{figure}

Our IFS data show a point-like feature south of T Tau N, which we identify as a tentative new component in the system. This new  companion is shown in Fig.~\ref{fig:ifsplots}, which shows
the PSF reference star TYC~1290-457-1, the T Tau system and 
the T Tau system with the PSF reference star subtracted at two different wavelengths (1.017 and 1.341 $\mu$m). 
The astrometric and photometric measurements of the new companion are computed with a 0.1 $\mu$m binning, and are listed in Table~\ref{table:obs}.

Although several artefacts can be seen in the IFS data shown in Fig.~\ref{fig:ifsplots}, the feature we label here as a tentative new companion is unique in several aspects: (1) the tentative companion is only seen in the T~Tau IFS data; no similar feature is present in the PSF data, (2) the tentative companion has the same position at the different wavelengths in the IFS data cube, excluding the possibility that this could be a speckle from the AO system, 
(3) the spectrum of the tentative component is much redder than the spectrum of T Tau N (the spectral slope in $F_\nu$ between 0.97--1.30 $\mu$m is 1.3 for the companion and 0.9 for T Tau N), excluding the possibility that this could be a reflection from T~Tau~N in the instrument's optics.

Close inspection of the IRDIS images in Fig.~\ref{fig:motions} also reveals a slightly extended point-like source just south of T Tau N, although its distance is close to the first Airy ring. As we do not have a PSF reference observation for the IRDIS frames, we cannot perform precise measurements in those frames and we rely on the IFS frames to determine the companion's properties.
The tentative companion is not detected in the IRDIS Ks images and is not found in the Kp- and H-band images made by  Keck/NIRC2 \citep{2014AJ....147..157S,2006A&A...457L...9D}. The IRDIS images have an  observational contrast limit of $\approx 3.3$ magnitude at 140~mas, one magnitude less than the brightness of the tentative companion, and the Keck/NIRC2 images have a contrast limit of $\approx 1.0$ at the same position. Therefore it is likely that the tentative companion is lost in the sky background in these images. 

The absolute magnitude of the tentative companion can be calculated using the 2MASS brightness of T Tau N, and the distance modulus derived by \citet{2005ApJ...619L.179L}, yielding $J_{mag}=6.0 \pm 0.4$. Assuming an age of the system of 2~Myr \citep{2001ApJ...556..265W}, and using the evolutionary tracks of \citet{1998A&A...337..403B}, this absolute magnitude is compatible with a stellar or sub-stellar companion with mass between 0.06 and 0.15~M$_\sun$.

\subsection{Astrometry}

\begin{figure*}
\centering
\includegraphics[width=0.49\hsize]{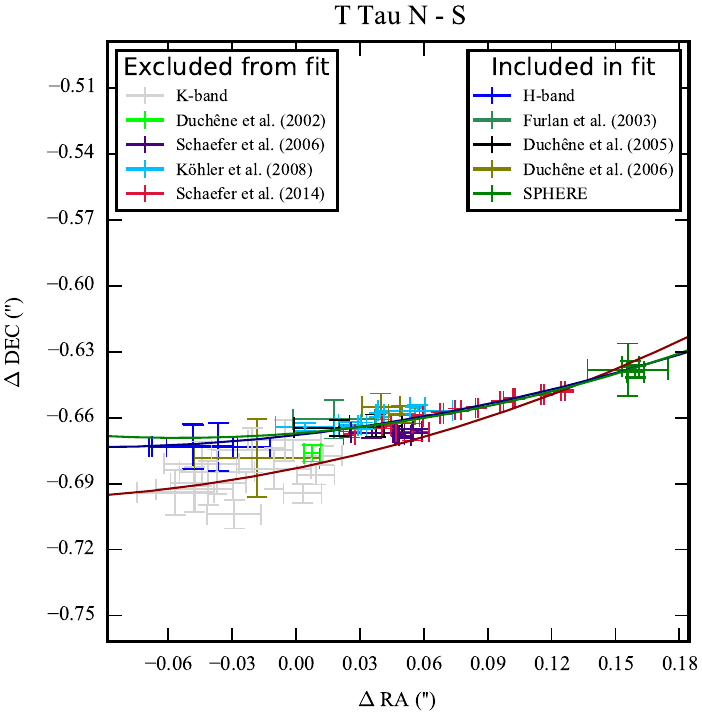}
\includegraphics[width=0.49\hsize]{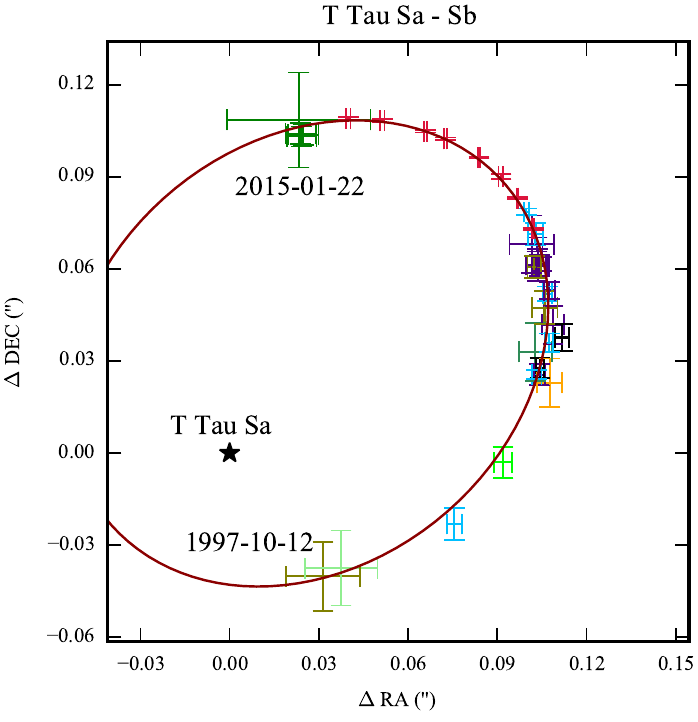}
\caption{\textit{Left:} Epochs of T~Tau~S around T~Tau~N. The coordinates are relative to T~Tau~N. 
Observations in which the Sa/Sb pair is not resolved are shown in either light gray or blue. The coloured markers show the resolved observations, in which the center of mass was calculated from the position of Sa and Sb. The red line shows the orbit calculated by \citet{2008A&A...482..929K}. 
The dark blue and green lines shows two possible orbital fits with the extreme periods of $2.7 \times 10^4$ and 475 years, respectively.
\textit{Right:} Epochs of T~Tau~Sb around T~Tau~Sa. The coordinates are relative to T~Tau~Sa.  The red line shows the orbit calculated by \citet{2014AJ....147..157S}.
}
\label{fig:ttaun_s}
\end{figure*}

We compared our new astrometric measurements of the relative positions of T~Tau~N, Sa and Sb 
with all available astrometry of the T Tau system from the literature 
(Fig.~\ref{fig:ttaun_s}). 
The literature contains (mostly recent) astrometric measurements in which Sa and Sb have been spatially 
resolved, and (mainly older) measurements in which only one southern component was detected. 
In our IFS observations T Tau Sa becomes visible at wavelengths longer than 1.4~$\mu$m, consistent with 
the IRDIS images, where T Tau Sa is only visible in the longer wavelength observations. Therefore we can 
safely assume that any observation at wavelengths shorter than 1.4~$\mu$m that  
did not resolve the southern components, has in fact observed T Tau Sb. We thus interpret those measurements 
as reflecting the position of T Tau Sb. We also know that the luminosity of T Tau Sa is comparable to 
T Tau Sb in the K-band and highly variable \citep{2010A&A...517A..16V, 2014AJ....147..157S}. This makes 
it impossible to use unresolved K-band observations to determine the exact position of Sa and Sb. We 
therefore omit from our analysis any literature observations in the K-band, in which Sa and Sb have not been resolved. 

As can be seen from Fig.~\ref{fig:ttaun_s}, significant orbital motion of the Sa-Sb 
binary around T~Tau~N, and in the Sa-Sb system itself has been detected. The positions are relative to either T~Tau~N (left) or T~Tau~Sa (right). 
The observations are labelled after the literature source. The unresolved observations are shown in light grey for the K-band, and blue for the H-band. Both are corrected for the motion of T Tau Sb around T Tau Sa, using the orbital parameters from \citet{2014AJ....147..157S}. It can be seen that the H-band data align with the resolved observations, whereas the K-band data (which are affected by the variable brightness ratio of Sa/Sb) seem to deviate. 

The corrected H-band points and the epochs from resolved observations miss the orbital fit of the T~Tau~S  around T~Tau~N (the center of gravity of Sa and Sb) from \citet{2008A&A...482..929K}, clearly showing that this previously derived orbit does not correctly take into account the orbital motion of T Tau Sb around Sa.
Therefore we searched for new orbital fits. 
We have excluded some observations from the orbital fit as their uncertainties were too low to acquire a $\chi_\nu^2<1$ fit, as shown in Fig.~\ref{fig:ttaun_s}.
There are also radio observations of this system, which list astrometric positions for presumably T~Tau~Sb, see \citet{2003AJ....125..858J} and \citet{2003A&A...406..957S}.
However, even after correcting for the orbital motion of Sb around Sa, the radio observations
do not align with the optical observations, and rather group with the unresolved K-band observations. 
This suggests that the origin of the radio emission may be more complex, and it may not originate only from T~Tau~Sb, therefore we omitted them from the plot and the orbital fit.

A random grid search of the parameter space yields a wide range of possible solutions for the orbit of T Tau S around T Tau N. The possible range of orbital parameters is shown in Table~\ref{table:orbpars}.
We note that some of the parameters listed in this table exclude the orbital solution derived by \citet{2008A&A...482..929K}. This is a natural consequence of the clear differences in our fits as can seen from Figure~\ref{fig:ttaun_s}.
We note that, assuming the same measurement precision that we achieved with SPHERE, a unique solution for the orbit should be possible with new data obtained around 2027--28.
Regarding the orbital motion of Sa around Sb, our fit is essentially identical to the orbital fit by \citet{2014AJ....147..157S}.

\begin{table}
\setlength{\tabcolsep}{2.5pt}
\caption{\label{table:orbpars}New orbital parameters for T Tau N-S.}
\centering
\begin{tabular}{lrcl}
\hline\hline
Parameter & \multicolumn{3}{c}{Range} \\
\hline
a (\arcsec)                       & $0.57$ & $-$ & $16.8$   \\
a (AU)                            & $84 $ & $-$ & $ 2465$ \\
Eccentricity                      & $0.10 $ & $-$ & $ 0.98$  \\
Inclination ($^\circ$)            & $0.6 $ & $-$ & $ 78.1$  \\
$\omega$ ($^\circ$)               & $1.2 $ & $-$ & $ 178.0$ \\
$\Omega$ ($^\circ$)               & $14.1 $ & $-$ & $ 176.7$ \\
Period (years)                    & $475 $ & $-$ & $ 27000 $ \\
T$_{\mathrm{periastron}}$ (years) & $1400 $ & $-$ & $ 7200$ \\
\hline
\end{tabular}
\end{table}

\section{Summary and Conclusions}

In the new SPHERE data on the T Tauri system presented here, two components, T~Tau~N and Sb are 
visible in all classical images, while Sa only becomes visible at wavelengths longer than 
1.4~$\mu$m. All components appear indistinguishable from a point-source. 
We detected a new tentative companion south of T Tau N, at a position of $144 \pm 3$~mas, PA: $197 \pm 2^\circ$, relative to T Tau N. This companion is not present in the PSF reference observation, and its spectrum differ from the other stellar companions in the system. If this is a new stellar or sub-stellar source within the system, we estimate its mass to be between 0.06 and 0.15~M$_\sun$.

\nocite{2000ApJ...531L.147K} 
\nocite{2002ApJ...568..771D}
\nocite{2003ApJ...596L..87F}
\nocite{2003AJ....125..858J}
\nocite{2003A&A...406..957S}
\nocite{2004ApJ...614..235B}
\nocite{2005ApJ...628..832D}
\nocite{2006AJ....132.2618S}
\nocite{2006A&A...457L...9D}
\nocite{2008A&A...482..929K}
\nocite{2014AJ....147..157S}

New astrometric measurements of the relative positions of T~Tau~N, Sa and Sb were compared with 
astrometric data from the literature. The astrometric epochs agree with the orbital fit 
from \citet{2014AJ....147..157S} for the expected orbital motion of Sb around Sa. 
The orbit of the T~Tau~S system around T~Tau~N is poorly constrained by the existing observations
and can be fit with a range of orbits ranging from a nearly circular orbit with a period of 475 years to highly eccentric orbits with periods up to $2.7 \times 10^4$ years.

T~Tau~Sa appeared 0.92 magnitude brighter than T~Tau~Sb in the CntK2 image, which is in contrast to the latest $K$-band brightness ratio measurements from January 2014 \citep{2014AJ....147..157S}, where Sb was brighter than Sa. This photometric variability is similar to the behaviour shown prior to 2006 \citep{2004ApJ...614..235B, 2005ApJ...628..832D}, 
and thus demonstrates that the period of lesser variability, when Sa and Sb had comparable brightnesses between 2006--2013, has come to an end. 
The brightness ratio H$_2$/CntK2 is 1.90, very high for T~Tau~N, demonstrating the presence of strong H$_2$ 1-0 S(1) emission close to this component. 
Previous studies found H$_2$ emission around T~Tau~N and S \citet{2010A&A...517A..19G}, which our observations also confirm.

\begin{acknowledgements} 
The authors would like to thank the SPHERE Science Verification team for their support in obtaining 
the observations.

This work was supported by the Momentum grant of the MTA CSFK Lend\"ulet Disk Research Group, and the Hungarian Research Fund OTKA grant K101393. 
\end{acknowledgements}

\bibliographystyle{aa}
\bibliography{ref_ttau_v5}

\Online 

\onlfig{
\begin{figure}
\centering
\includegraphics[width=0.85\hsize]{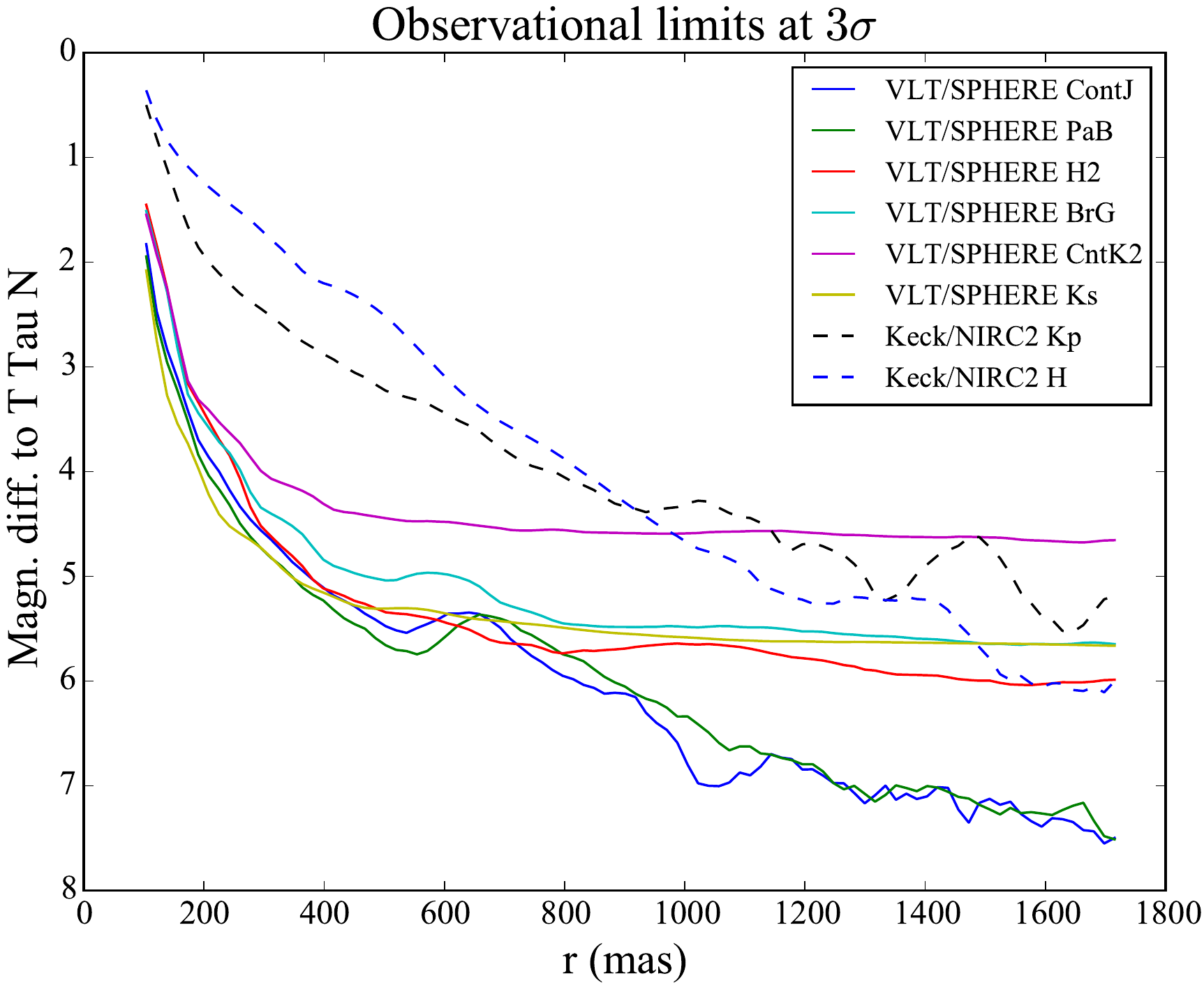}
\caption{The 3-$\sigma$ detection limits for point-like sources near T~Tau N in the classical imaging IRDIS mode of VLT/SPHERE and Keck/NIRC2. The bumps in the SPHERE data at $\approx 250$ and $\approx 600$ mas are the wings of the PSF from T Tau N. The Keck/NIRC2 data is obtained from the Keck Observatory Archive, from projects identified as N26N2 for the H-band and 2010B for the K-band data. 
}
\label{fig:obslimits}
\end{figure}
}

\onlfig{
\begin{figure}
\centering
\includegraphics[width=0.60\hsize]{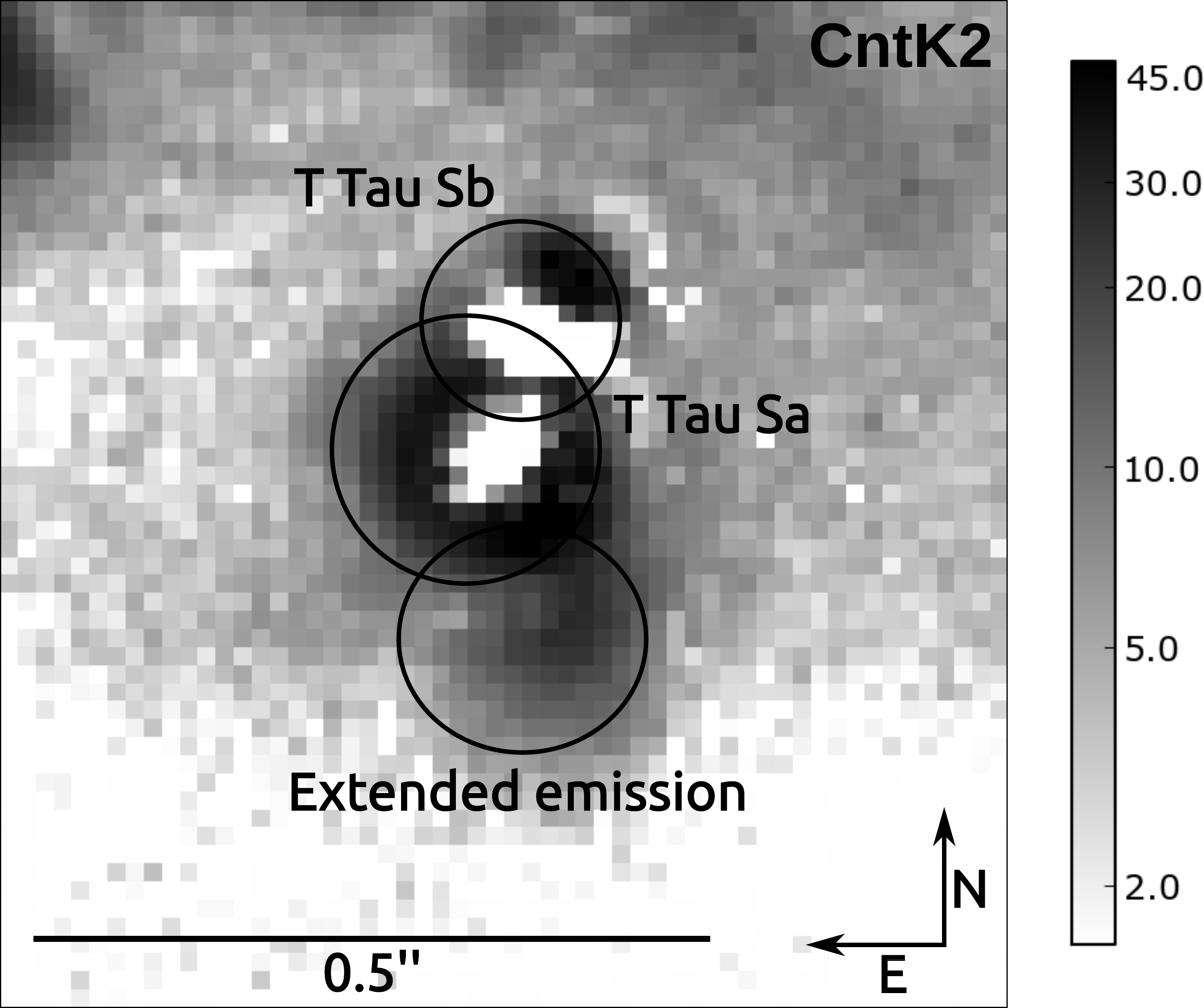}
\caption{Extended emission south of T Tau Sa. Both T Tau Sa and Sb were subtracted using T Tau N as a PSF reference scaled to the peak of T Tau Sa and Sb, respectively. 
The leftover artefacts are due to the Airy rings apparent around all three components, 
which cannot be totally eliminated because of the closeness of T Tau Sa and Sb. 
The leftover residuals have an amplitude of 2.4\% and 6.6\% of the peak fluxes of T Tau Sa and Sb, respectively.
}
\label{fig:cntk2_subtr}
\end{figure}
}

\end{document}